\documentclass[aps,prl,preprint,groupedaddress,showpacs]{revtex4}
\usepackage{graphicx}
\begin{document}

\title{Comment on 'Pulsed field studies of the magnetization reversal in molecular nanomagnets'}

\author{W. Wernsdorfer$^1$, N. E. Chakov$^2$, and G. Christou$^2$}

\affiliation{
$^1$Laboratoire Louis N\'eel, associ\'e \`a l'UJF, CNRS, BP 166, 38042 Grenoble Cedex 9, France\\
$^2$Department of Chemistry, University of Florida, Gainsville, FL 
32611-7200, US
}

\date{30 April 2004}

\begin{abstract}
In a recent paper, http://xxx.lanl.gov/abs/cond-mat/0404041,
J. Vanacken et al. reported experimental studies of crystals of 
Mn$_{12}$-ac molecular nanomagnets in pulsed magnetic fields with sweep rates up to
4000 T/s. Steps in the magnetization curve were observed. 
The data were explained by collective dipolar relaxation. 
We give here an alternative
explanation that is based on thermal avalanches triggered
by {\it defect} molecules (faster relaxing species). 
These species are always present in Mn$_{12}$-ac molecular 
nanomagnets. We propose a simple method to test this interpretation.
Note also that we do not question the possibility
of collective effects that are bassed on spin--spin
interactions.
\end{abstract}

\maketitle

In a recent paper, http://xxx.lanl.gov/abs/cond-mat/0404041,
J. Vanacken et al. reported low temperature magnetization
studies of Mn$_{12}$-ac single crystals at a field sweep rate
up to 4000 T/s~\cite{Vanacken}. Their main finding was that at such
high sweep rates the position of the steps in the magnetic
relaxation shifts by $\Delta H$ that increases as the sweep rate
goes up. The authors scaled the relaxation curves
obtained at different sweep rates onto one curve. The
scaling was explained within a model of collective
magnetic relaxation of the crystal.

Instead of commenting on the interpretations 
by J. Vanacken et al.~\cite{Vanacken},
we point out here important facts about Mn$_{12}$-ac 
that may have disrupted their experiments and propose
an alternative explanation that is based on thermal 
avalanches triggered
by {\it defect} molecules (faster relaxing species). 
These species are always present in Mn$_{12}$-ac molecular nanomagnets.
We then propose a simple method to test this interpretation.
Note also that we do not question the possibility
of collective effects that are bassed on spin--spin
interactions~\cite{WW_PRL02}. Such effects can lead
to 'off resonance' tunneling in the case of 
sufficient energy level mixing (activated levels, high fields, etc.).

\begin{figure}
\begin{center}
\includegraphics[width=.8\textwidth]{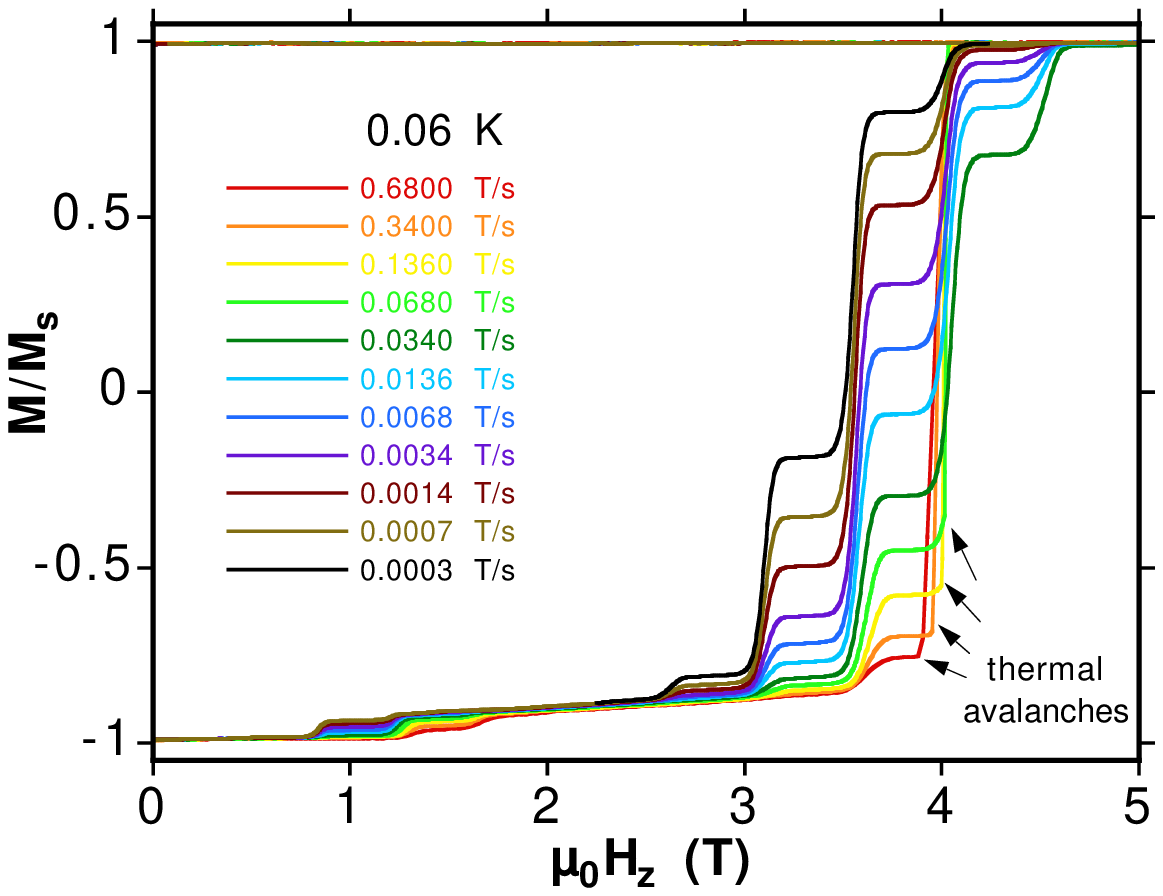}
\includegraphics[width=.8\textwidth]{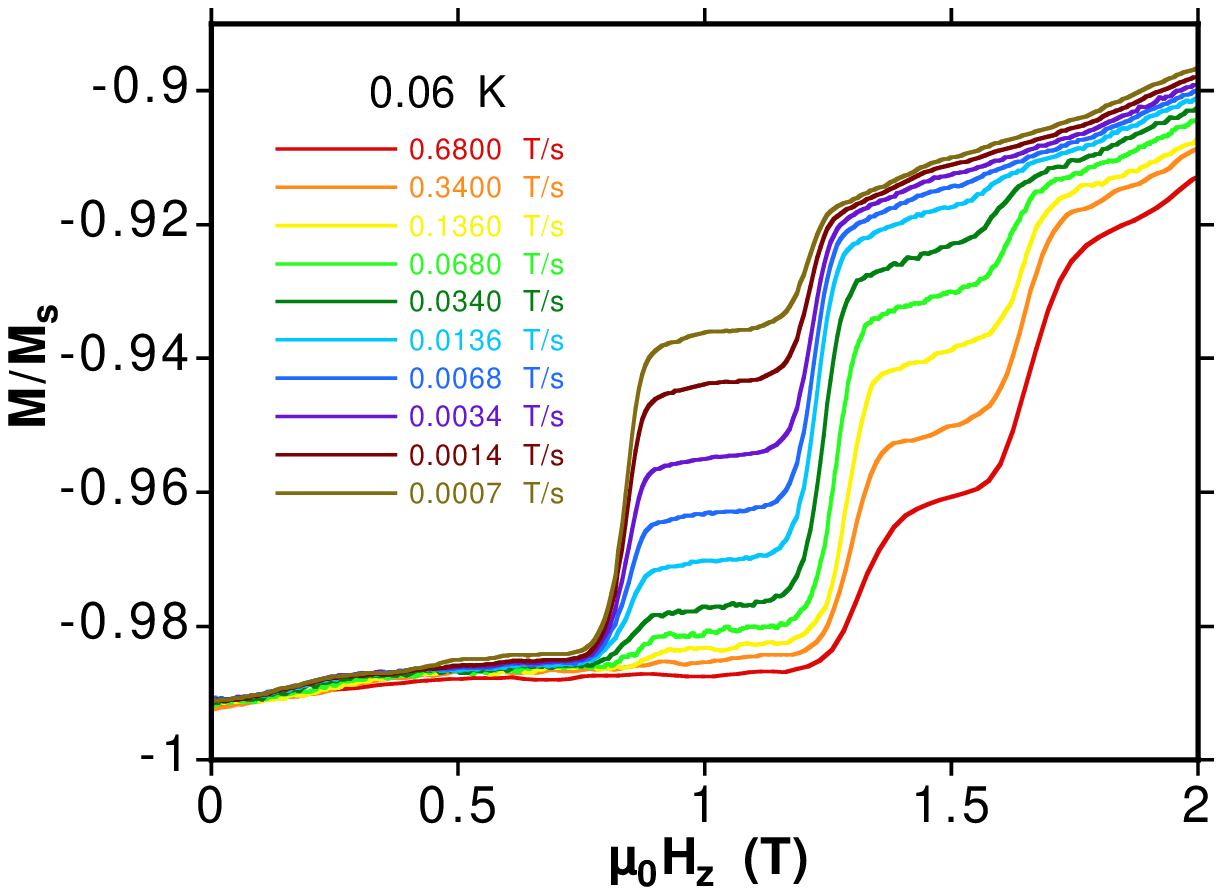}
\caption{(color)  (a) Hysteresis loop measurements
for a single crystal of Mn$_{12}$-ac at several
field sweep rates. The field was applied at an
angle of about 6$^{\circ}$ in respect to the 
easy axis of magnetization. (b) Enlargement 
for the low field region showing the relaxation
of the fast species.}
\label{fig2}
\end{center}
\end{figure}

Fig. 1 shows typical hysteresis loops~\cite{noteHall} 
for a single crystal of Mn$_{12}$-ac.
The cryostat temperature was about 60 mK.
When the applied field is near an avoided level 
crossing, the magnetization relaxes faster, 
yielding steps separated by plateaus~\cite{Kent00,ChiorescuMn12PRL00}. 
The loops show two series of steps: in
the low field region (0 to 2 T) and in the
high field region (2.5 to 5 T). It is now
well known~\cite{remark1} that the steps in the
low field region are due to faster relaxing species
whereas the others are due
to the {\it normal} species of 
Mn$_{12}$-ac~\cite{Kent00,ChiorescuMn12PRL00}.
The step heights decrease for faster field 
sweep rates. However, concerning the four fastest 
field sweep rates, a thermal avalanche~\cite{avalanches} 
is observed at about 4 T that reverses the entire
magnetization in a time scale faster than a millisecond (the time
resolution of the lookin amplifier). The avalanche field
decreases for faster sweep rates but the avalanche
is always observed close to a resonance field.

When comparing the measurements in Fig. 1 with the 
pulsed field ones of
J. Vanacken et al. (Fig. 2 in~\cite{Vanacken}), we note
that their field sweep rates are
more than 1000 times larger than our rates.
Furthermore, our crystal size is much smaller than
in~\cite{Vanacken}. Both points should enhance
the thermal avalanche effect in the pulsed field experiment.
We also note in the experiment by Vanacken et al. 
that the entire magnetization reverses
in a time scale of about 0.2 ms and in the low field
region (below 2 T) (see Fig. 2 in~\cite{Vanacken}). 
We propose therefore that this
fast magnetization reversal could have been triggered by the heat
emission of the fast relaxing species reversing below
2 T~\cite{note_fin}. The field sweep rate dependence of the step position
could reflect the time needed to trigger the thermal avalanche.

This interpretation can be checked with the following
method. First, a high negative field should be applied to saturate
the magnetization. Then, the field should be swept slowly up to 2 T
reversing only the fast relaxing species. Finally, the field
should be ramped to zero field and a positive field pulse applied.
The magnetization reversal should then occur above 3 T because
the fast relaxing species cannot trigger any more a thermal
avalanche. Note that all relevant tunnel splittings 
of the {\it normal} species are too small to
induce a significant reversal below 3 T and below 0.6 K.
We do not expect therefore an thermal avalanche below
about 3 T.



\begin{thebibliography}{1}
\bibitem{noteHall}
All measurements were performed using a 
2D electron gas micro-Hall probe. 
The high sensitivity  allows the study of single 
crystals of SMMs of the order of 10 to 500 $\mu$m. 
The sample of the present study was $20 \times 6 \times 5$ $\mu$m$^3$.
The field can be applied in any direction by separately 
driving three orthogonal coils.
\bibitem{remark1}  
Several authors have pointed out that in the Mn$_{12}$ carboxylate family
different isomeric forms give rise to different relaxation rates.
This was first observed in Mn$_{12}$-ac 
\cite{Ruiz98,Aubin98,Sun98,Sun99}
and has been studied in detail \cite{Aubin01,Aubin01b,Soler03,Soler03b}. 
We found that a minor species of Mn$_{12}$-ac \cite{WW_EPL99}, randomly
distributed in the crystal, exhibits a faster
relaxation rate which becomes temperature independent below 0.3 K.
\end{thebibliography}

\end{document}